\documentclass[twocolumn,english,aps,prd,reprint,floatfix,notitlepage,footinbib,preprintnumbers,superscriptaddress]{revtex4-1}
\pdfoutput=1
\usepackage{lmodern}

\usepackage[T1]{fontenc}
\usepackage[latin9]{inputenc}
\usepackage{geometry}
\usepackage{subfigure,lmodern, amsmath,amssymb, graphicx, pifont, adjustbox, bm, xcolor}
\usepackage{amsfonts}
\usepackage{geometry}
\usepackage{comment}
\usepackage{mathtools}
\usepackage{float}
\usepackage{slashed}
\usepackage{ragged2e}
\usepackage{array}
\usepackage{hhline}

\makeatletter\g@addto@macro\bfseries{\boldmath}\makeatother

\newcommand{\SigmaScaled}{\scalebox{1.3}{\raisebox{-0.35mm}{$\Sigma$}}}

\usepackage{stackengine}
\usepackage{esint}
\usepackage[unicode=true,pdfusetitle,
 bookmarks=true,bookmarksnumbered=false,bookmarksopen=false,
 breaklinks=false,pdfborder={0 0 1},backref=false,colorlinks=true]
 {hyperref}
\hypersetup{
 pdfauthor={Clifford Cheung, Aaron Hillman, and Grant N. Remmen},
 citecolor=black,linkcolor=black,urlcolor=black}

\newcommand{\appendixref}[1]{\hyperref[#1]{appendix~\ref{#1}}}
\def\equationautorefname~#1\null{eq.\,(#1)\null}
\usepackage{breakurl}
\usepackage{breakurl}
\usepackage[hang,flushmargin]{footmisc} 
\allowdisplaybreaks
\makeatletter

\usepackage{etoolbox}
\apptocmd{\thebibliography}{\justifying\setlength{\leftskip}{7.4mm}}{}{} 
 
 \usepackage{relsize}
\usepackage{babel}

\makeatletter
\def\simgt{\mathrel{\lower2.5pt\vbox{\lineskip=0pt\baselineskip=0pt
           \hbox{$>$}\hbox{$\sim$}}}}
\def\simlt{\mathrel{\lower2.5pt\vbox{\lineskip=0pt\baselineskip=0pt
           \hbox{$<$}\hbox{$\sim$}}}}
\makeatother

\usepackage{changepage}

\newcommand{\be}{\begin{equation}}
\newcommand{\ee}{\end{equation}}
\newcommand{\bea}{\begin{eqnarray}}
\newcommand{\eea}{\end{eqnarray}}

\newcommand{\eq}[2]{\be\begin{aligned}#1 \label{#2}\end{aligned}\ee}

\newcommand{\mysec}[1]{\noindent {\bf #1.}---}
\newcommand{\mysubsec}[1]{\noindent {\it #1.}---}

\newcolumntype{P}[1]{>{\centering\arraybackslash}p{#1}}

	\definecolor{dartmouthgreen}{rgb}{0.05, 0.5, 0.06}

\begin{document}

\preprint{CALT-TH 2024-030}

\title{Uniqueness Criteria for the Virasoro-Shapiro Amplitude}

\author{Clifford Cheung}
\affiliation{Walter Burke Institute for Theoretical Physics, California Institute of Technology, Pasadena, CA 91125}
\author{Aaron Hillman}
\affiliation{Walter Burke Institute for Theoretical Physics, California Institute of Technology, Pasadena, CA 91125}
\author{Grant N. Remmen}
\affiliation{Center for Cosmology and Particle Physics, Department of Physics, New York University, New York, NY 10003}
    
\begin{abstract}

\noindent 
The scattering amplitudes of string theory exhibit many extraordinary properties. But are they the unique mathematical objects to do so?  Recently, it has been shown how the spectrum and amplitudes of open string theory follow directly from the assumptions of faster than power-law falloff at high energies and a property dubbed level truncation.  At present there is no analogous principle for closed string scattering, which is famously rigid and naively impervious to modification.  In this paper we analytically bootstrap the spectrum and four-point amplitudes of the closed string---together with a parameterized space of deformations---from conditions on high-energy falloff and level truncation. While these deformations exhibit the same Regge scaling as pure gravity, in the tensionless limit they reproduce remarkable extremal amplitudes that have appeared in bottom-up studies of positivity. 
\end{abstract}
\maketitle

\mysec{Introduction}Is string theory the only game in town?  This question has generated much debate, but it is rarely interpreted as a bona fide scientific hypothesis to be evaluated as either true or false.  As it turns out, there is an approach that can attack this question with  mathematical precision, which is known as the bootstrap.  At a very basic level, the bootstrap works by constructing a function space for all possible observables and then constraining it with physical principles.   This is an inversion of the usual top-down logic: usually, one assumes a string theoretic framework, which then implies scattering amplitudes that exhibit numerous miraculous features like tame high-energy behavior, dual resonance, and more.  Instead, the bootstrap takes these features as input assumptions and asks whether string amplitudes are the unique solution to this set of constraints.


The approach of the string bootstrap is very old but has enjoyed a resurgence of activity and genuine progress.  
Historically, essentially all work on this subject has been devoted to the  Veneziano amplitude~\cite{Veneziano} of open string theory and its deformations.   
While the seminal papers on this topic trace back more than half a century~\cite{Coon:1969yw,Gross:1969db,Mandelstam:1968czc}, recent efforts~\cite{Cheung:2022mkw,Cheung:2023adk,Cheung:2023uwn,Haring:2023zwu,Bhardwaj:2024klc,Maldacena:2022ckr,Bjerrum-Bohr:2024wyw,Bhardwaj:2023eus,Eckner:2024ggx,Eckner:2024pqt,Geiser:2022icl,Geiser:2022exp,Figueroa:2022onw,Chakravarty:2022vrp,Geiser:2023qqq,Jepsen:2023sia,Bhardwaj:2022lbz} have  identified a collection of Veneziano variations that reproduce many of its miraculous mathematical properties.

Ironically, even though the raison d'\^{e}tre of string theory is to provide a theory of quantum gravity, much less is understood about whether closed string amplitudes are unique, or how to bootstrap them analytically.  The Virasoro-Shapiro (VS) amplitude~\cite{Virasoro,Shapiro},
\eq{
M_{\rm VS}(s,t) = - \frac{\Gamma(-s)\Gamma(-t)\Gamma(-u)}{\Gamma(1+s)\Gamma(1+t)\Gamma(1+u)},
}{eq:MVS} 
is the only definitively consistent ultraviolet completion of graviton scattering \footnote{Here we have stripped off the prefactor involving graviton polarizations and have chosen convenient units for the string tension.}.  Furthermore, it is very resistant to deformation~\cite{Cheung:2023adk,Geiser:2022icl,Geiser:2022exp,Chowdhury:2019kaq}, though numerical approaches to constructing graviton amplitudes have been fruitful~\cite{Guerrieri:2021ivu,Caron-Huot:2021rmr,Bern:2021ppb}.  In the analytic approach, however, the known deformations of VS either violate locality~\cite{Huang:2022mdb} or are constructed in terms of artificial ``satellite amplitudes,'' which are literally just sums over the original VS amplitude with shifted kinematic invariants~\cite{Arkani-Hamed:2020blm,Cheung:2023adk,Albert:2024yap}.
To date, there is no bootstrap principle that elevates VS above these satellite amplitudes.
Consequently, identifying {\it any set of bootstrap principles} that singles out the VS amplitude cuts to the heart of the question of whether string theory is the unique theory of quantum gravity.

Recent work~\cite{Cheung:2024uhn} has identified a simple set of bootstrap principles that uniquely fix the Veneziano amplitude {\it without assuming} the spectrum of masses~\footnote{In contrast, Ref.~\cite{Caron-Huot:2016icg} showed that the open string amplitude was {\it asymptotically} unique under broad technical assumptions that have recently been shown to be violated in certain cases~\cite{Haring:2023zwu}. Alternatively, adding the strong assumption of string monodromy allows for the bootstrapping of Veneziano from dispersion relations~\cite{Berman:2023jys,Chiang:2023quf}. Historically, Coon~\cite{Coon:1969yw} bootstrapped the spectrum of his $q$-deformation of the Veneziano amplitude by requiring infinite product form of the bootstrap.}.  
The assumptions of this bootstrap are dual resonance---i.e., a vanishing amplitude at high energies---and {\it level truncation}, which posits a sequence of values of the exchanged momentum at which the amplitude reduces to a rational function in the center-of-mass energy. 
Dual resonance, level truncation, planar crossing symmetry, and superpolynomial softness~\cite{Haring:2023zwu} were then sufficient to uniquely construct the open string amplitude in Ref.~\cite{Cheung:2024uhn}.
Relaxing the final condition yielded old and new deformations of Veneziano.
Remarkably, this bootstrap is analytically solved without assuming the locations of mass poles and residue zeros, which are instead taken to be arbitrary. Unlike all previous analyses, the string spectrum is an output rather than an input of the bootstrap. 

In this paper, we apply level truncation to the qualitatively different problem of bootstrapping the amplitude of the {\it closed string}.
Our starting point is a tree-level~\footnote{Straightforward $\hbar$ counting is sufficient to show that any perturbative ultraviolet completion of gravity must be implemented by tree-level exchanges~\cite{Cheung:2016wjt}.}, fully permutation symmetric---i.e., nonplanar---amplitude ansatz exhibiting dual resonance, along with the appropriate generalization of level truncation.
Again, the spectrum is not assumed.
Solving this bootstrap problem yields a two-parameter family of deformations of the VS amplitude.  Notably, an output of the bootstrap is that the spectrum of massive states must be linear.
After characterizing the ultraviolet behavior, we find that in certain parameter limits this bootstrapped amplitude exactly reproduces  extremal corners of the consistent space of couplings for the effective field theory (EFT) dictated by unitarity, analyticity, and causality~\cite{Caron-Huot:2020cmc}.
We find regions of parameter space consistent with partial wave unitarity and construct planar analogues of these amplitudes.
Finally, we show that invoking superpolynomial softness in addition to level truncation {\it uniquely} yields the VS amplitude.

\medskip

\mysec{Amplitude Bootstrap}Throughout, we use mostly-plus metric signature and define the Mandelstam variables as $s=-(p_1+p_2)^2$, $t = -(p_2+p_3)^2$, and $u=-(p_1+p_3)^2$, where the $p_i$ are the cyclically labeled external momenta.
For an amplitude that vanishes in the Regge limit $\lim_{s\rightarrow \infty} M(s,t) = 0$
for some values of $t$, analyticity implies the existence of a dual resonant representation,
\be
M(s,t)=\sum_{n=0}^{\infty}\left(\frac{1}{\mu(n)-s}+\frac{1}{\mu(n)-u}\right)R(n,t),\label{eq:DR}
\ee
where we allow poles in both the $s$ and $u$ channels.
While symmetry in $s\leftrightarrow u$ is manifest in the ansatz, for full permutation invariance---and therefore $t$ channel poles---we must additionally enforce $s\leftrightarrow t$ crossing, which will nontrivially constrain the residue polynomials $R(n,t)$.

\medskip

\mysubsec{Level Truncation}The residues of the VS amplitude are
\eq{
R_{\rm VS}(n,t) \,{=} \!\left(\!\frac{\Gamma(t+n)}{\Gamma(1{+}n)\Gamma(1{+}t)}\!\right)^2 \!{=}\, \frac{1}{(n!)^2} \prod_{k=1}^{n-1}(t + k)^2
}{}
for massive states with $n\geq 1$, and they plainly exhibit level truncation: when $t=-k$, $R_{\rm VS}(n,t)\,{=}\,0$ for all positive integers $n\,{>}\,k$.
When level truncation occurs, the dual resonant sum terminates at a finite number of terms, and the amplitude becomes a rational polynomial in $s$.
Note, however, that $\partial_t M_{\rm VS}(s,t)$ {\it also exhibits level truncation} at $t=-k$, so $\partial_t R_{\rm VS}(n,t) = 0$ at $t=-k$ for all $n > k$.
In fact, the VS amplitude vanishes at $t=-k$, though we will not take this strong assumption as an input for our bootstrap.

Motivated by these observations, we take as a bootstrap assumption that both $M(s,t)$ and $\partial_t M(s,t)$ exhibit level truncation at an unknown nonrepeating sequence $t=\xi(k)$, at which they have poles in $s$  only up to level $n\,{=}\,k$~\footnote{A priori, one can make the more general assumption that $M(s,t)$ and $\partial_t M(s,t)$ each exhibit level truncation at distinct sequences, $t=\xi(k)$ and $t=\tilde \xi(k)$, respectively.  However, even in this case we can prove that enforcing these conditions implies that $\xi(k)=\tilde\xi(k)$.}.
Level truncation of $M$ and $\partial_t M$ then implies the following squared form for the residue polynomials,\vspace{-1mm}
\be
R(n,t) = c(n) \prod_{k=1}^{n-1} (t-\xi(k))^2, \label{eq:Ransatz}\vspace{-1mm}
\ee
which is highly suggestive of the double copy~\cite{Bern:2008qj} familiar from gauge theory and gravity.  To summarize, level truncation and vanishing Regge behavior imply a bootstrap ansatz given by Eqs.~\eqref{eq:DR} and \eqref{eq:Ransatz}, which  depends on three unknown infinite sequences: $\mu(n)$, $\xi(n)$, and $c(n)$.

\medskip

\mysubsec{Low-Energy Matching}To match onto low-energy gravity, we set $\mu(0)\,{=}\,0$ and $R(0,t)= \rho^2/t^2$ for the level $n\,{=}\,0$ state. We then have graviton exchange $M = \rho^2/stu + \cdots$, where $\rho$ parameterizes the strength of Newton's constant relative to the first heavy state coupling in units of $\mu(1)$.
We take massless external states, so $s+t+u=0$, choose mass units such that $\mu(1) = 1$, and scale the amplitude by a positive coupling to fix $c(1) = 1$.

\medskip

\mysubsec{Crossing Symmetry}At $(s,t) = (\xi(i),\xi(j))$, the dual resonant sum in Eq.~\eqref{eq:DR} truncates, with $n$ running from $0$ to $j$.
At these kinematics, crossing symmetry,
\be 
M(\xi(i),\xi(j)) = M(\xi(j),\xi(i)),\label{eq:discretecrossing}
\ee
yields an algebraic constraint on a finite set of variables among the $\mu(n)$, $\xi(n)$, and $c(n)$ for each choice of integers $(i,j)$.

These conditions define a complicated algebraic variety.
By expanding $c$, $\mu$, and $\xi$ in small perturbations about the VS amplitude, we can then solve the system of nonlinear algebraic equations at first order in perturbations, reducing the problem to a set of linear algebraic equations that are trivially solved.  This exercise yields a family of solutions parameterized by a single variable $\lambda$.  Remarkably, we find that these solutions, given by
\be 
\mu(n) = \tfrac{n+\lambda-1}{\lambda},\, \xi(n) =- \tfrac{2n+\lambda-1}{2\lambda},\,
c(n) = \tfrac{\lambda^{2n-2}}{\left(\frac{1+3\lambda}{2}\right)_{n-1}^2}\label{eq:sols}
\ee
for $n\geq 1$, where $(x)_n = \Gamma(x+n)/\Gamma(x)$ is the Pochhammer symbol, also solve the full nonlinear conditions in Eq.~\eqref{eq:discretecrossing}.  Furthermore, to verify that these solutions are unique, one can linearly perturb about the nonlinear solution above and verify that there are no additional solutions connected to it.
That is, Eq.~\eqref{eq:sols} is the {\it only} solution to the crossing equations that is continuously connected to VS, and it is likely the only solution in general. 
If we invoke the additional physical requirement that all odd angular momentum partial waves of the residue vanish, as required for the scattering of identical bosons, then we find that this solution is indeed unique.
Throughout, we require $\lambda \geq 0$ in order to avoid the spinning tachyons that arise for negative $\lambda$~\cite{Arkani-Hamed:2003pdi,Cheng:2006us,Dubovsky:2006vk}.

\medskip

\mysec{Amplitudes}Inputting the solutions in Eq.~\eqref{eq:sols} into our dual resonant ansatz in Eqs.~\eqref{eq:DR} and \eqref{eq:Ransatz}, we have the residues for $n\geq 1$,
\begin{equation}
R(n,t)=\left(\frac{\left(\frac{1+\lambda}{2}+\lambda t\right)_{n-1}}{\left(\frac{1+3\lambda}{2}\right)_{n-1}}\right)^{2}\label{eq:Resfinal}
\end{equation}
and a closed-form expression for the amplitude,
\be\hspace{-2mm}\vspace{2mm}
\begin{aligned}
&M(s,t) = \frac{\rho^2}{stu} \\&+\left(\frac{1}{1-s}\,{}_{4}F_{3}\left[\begin{smallmatrix}
1,\lambda(1-s),\frac{1+\lambda}{2}+\lambda t,\frac{1+\lambda}{2}+\lambda t\\
\frac{1+3\lambda}{2},\frac{1+3\lambda}{2},1+\lambda(1-s)
\end{smallmatrix};1\right] + s\leftrightarrow u\right).\hspace{-2mm}
\end{aligned} \label{eq:Mpre}
\ee
This amplitude can be put into an equivalent form that is manifestly permutation invariant,
\begin{widetext}
\be
M(s,t) = \frac{\rho^2}{stu}+\frac{\Gamma(3\lambda+1)\Gamma(\lambda-\lambda s)\Gamma(\lambda -\lambda t)\Gamma(\lambda -\lambda u)}{3\Gamma(2\lambda +\lambda s)\Gamma(2\lambda +\lambda t)\Gamma(2\lambda + \lambda u)}\,_{5}F_{4}\left[\begin{smallmatrix}
\lambda (1-s),\,\lambda (1-t),\,\lambda (1-u),\,3\lambda -1,\,\frac{3\lambda -1}{2}\\
\lambda (2+s),\,\lambda (2+t),\,\lambda (2+u),\,\frac{3\lambda+1}{2}
\end{smallmatrix};1\right].\label{eq:M}
\ee
\end{widetext}
When $\lambda = \rho = 1$, the amplitude reduces to the VS amplitude in Eq.~\eqref{eq:MVS}.
We can express Eq.~\eqref{eq:M} as an infinite satellite sum over affine transformed versions of the finite deformations of VS proposed in Refs.~\cite{Arkani-Hamed:2020blm,Cheung:2023adk},
\be
\begin{aligned}
M &= \frac{\rho^2}{stu} + \sum_{k=0}^{\infty}\frac{\lambda(3\lambda-1)^2 \Gamma(3\lambda-1+k)}{(3\lambda-1+2k)k!} M_k \\
M_k &=\frac{\Gamma(k{+}\lambda{-}\lambda s)\Gamma(k{+}\lambda {-}\lambda t)\Gamma(k{+}\lambda {-}\lambda u)}{\Gamma(k{+}2\lambda {+}\lambda s)\Gamma(k{+}2\lambda {+}\lambda t)\Gamma(k{+}2\lambda {+}\lambda u)}.
\end{aligned}\label{eq:Msatellite}
\ee
In fact, to derive Eq.~\eqref{eq:M} from Eq.~\eqref{eq:Mpre}, it is useful to first use Eq.~\eqref{eq:Msatellite}, which has the required residues in our solution in Eq.~\eqref{eq:Resfinal}, and which in turn can be straightforwardly written as Eq.~\eqref{eq:M} using the definition of the generalized hypergeometric functions.

\medskip

\mysubsec{Kinematic Limits}We write the EFT expansion of $M$ at low energies as
\be
\begin{aligned}
M =&\, \frac{\rho^2}{stu} + g_0 + g_2 (s^2 + t^2 + u^2)  \\&\hspace{6.38mm} + g_3 stu + g_4 (s^2 + t^2 + u^2)^2 + \cdots. 
\end{aligned}
\ee
We can extract $g_{0}$ from the $O(s^{0})$ term in the expansion
of the forward amplitude. Referring to the dual resonant form of the amplitude in Eq.~\eqref{eq:DR} and the residues in Eq.~\eqref{eq:Resfinal}, we observe that when $t=0$, we can perform the expansion in $s$ simply by expanding the propagators, from which
we obtain
\begin{equation}
g_{0}=2\sum_{n=1}^{\infty}\frac{R(n,0)}{\mu(n)}.
\end{equation}
From the definition of the generalized hypergeometric functions, we immediately recognize
\begin{equation}
g_{0}=2\,_{4}F_{3}\left[\begin{smallmatrix}1,\lambda,\frac{1+\lambda}{2},\frac{1+\lambda}{2}\\
1+\lambda,\frac{1+3\lambda}{2},\frac{1+3\lambda}{2}
\end{smallmatrix};1\right].
\end{equation}
Similarly extracting the $O(s^{2})$ term in the propagator expansion
of the forward amplitude in Eq.~\eqref{eq:DR}, we obtain $g_{2}$,
\begin{equation}
g_{2}=\sum_{n=1}^{\infty}\frac{R(n,0)}{\mu(n)^{3}}.
\end{equation}
Again from the definition of the generalized hypergeometric functions,
we have
\begin{equation}
g_{2}=\,_{6}F_{5}\left[\begin{smallmatrix}1,\lambda,\lambda,\lambda,\frac{1+\lambda}{2},\frac{1+\lambda}{2}\\
1+\lambda,1+\lambda,1+\lambda,\frac{1+3\lambda}{2},\frac{1+3\lambda}{2}
\end{smallmatrix};1\right].
\end{equation}
We can analogously extract $g_{4}$ from the $O(s^{4})$ term in the
forward amplitude in the dual resonant form of $M$,
\begin{equation}
g_{4}=\frac{1}{2}\sum_{n=1}^{\infty}\frac{R(n,0)}{\mu(n)^{5}}.
\end{equation}
Once more from the definition of the generalized hypergeometric functions,
we immediately recognize
\begin{equation}
g_{4}=\frac{1}{2}\,_{8}F_{7}\left[\begin{smallmatrix}1,\lambda,\lambda,\lambda,\lambda,\lambda,\frac{1+\lambda}{2},\frac{1+\lambda}{2}\\
1+\lambda,1+\lambda,1+\lambda,1+\lambda,1+\lambda,\frac{1+3\lambda}{2},\frac{1+3\lambda}{2}
\end{smallmatrix};1\right].
\end{equation}
Extracting $g_{3}$ is more complicated. From the dual resonant form
of the amplitude, we see that
\begin{equation}
\begin{aligned}
g_{3}&=-\frac{1}{2}\lim_{s,t\rightarrow0}\partial_{t}\partial_{s}^{2}\left(M(s,t)-\frac{\rho^{2}}{stu}\right)\\&=\sum_{n=1}^{\infty}\frac{3}{\mu(n)^{4}}R(n,0)-\lim_{t\rightarrow0}\frac{{\rm d}}{{\rm d}t}\sum_{n=1}^{\infty}\frac{2}{\mu(n)^{3}}R(n,t).
\end{aligned}
\end{equation}
That is,
\begin{equation}
\begin{aligned}
g_{3}&= -2\lim_{t\rightarrow0}\frac{{\rm d}}{{\rm d}t}\sum_{n=0}^{\infty}\frac{(\lambda)_{n}^{3}}{(1+\lambda)_{n}^{3}\left(\frac{1+3\lambda}{2}\right)_{n}^{2}}\left(\tfrac{1+\lambda}{2}+\lambda t\right)_{n}^{2}\\&\qquad +3\sum_{n=0}^{\infty}\frac{(\lambda)_{n}^{4}\left(\frac{1+\lambda}{2}\right)_{n}^{2}}{(1+\lambda)_{n}^{4}\left(\frac{1+3\lambda}{2}\right)_{n}^{2}}.
\end{aligned}
\end{equation}
Defining $\nu(t)=\lambda+2\lambda t$, we see that $\lim_{t\rightarrow0}\frac{{\rm d}}{{\rm d}t}=2\lim_{\nu\rightarrow\lambda}\frac{{\rm d}}{{\rm d}\log\nu}$,
and so
\begin{equation}
\begin{aligned}
g_{3}&=-4\left.\frac{{\rm d}}{{\rm d}\log\nu}\sum_{n=0}^{\infty}\frac{(\lambda)_{n}^{3}\left(\frac{1+\nu}{2}\right)_{n}^{2}}{(1+\lambda)_{n}^{3}\left(\frac{1+3\lambda}{2}\right)_{n}^{2}}\right|_{\nu=\lambda}\\&\qquad +3\sum_{n=0}^{\infty}\frac{(\lambda)_{n}^{4}\left(\frac{1+\lambda}{2}\right)_{n}^{2}}{(1+\lambda)_{n}^{4}\left(\frac{1+3\lambda}{2}\right)_{n}^{2}}.
\end{aligned}
\end{equation}
From the definition of the generalized hypergeometric functions, we
at last have
\begin{equation}
\begin{aligned}
g_{3}&=\left.-4\frac{{\rm d}}{{\rm d}\log\nu}\,_{6}F_{5}\left[\begin{smallmatrix}1,\lambda,\lambda,\lambda,\frac{1+\nu}{2},\frac{1+\nu}{2}\\
1+\lambda,1+\lambda,1+\lambda,\frac{1+3\lambda}{2},\frac{1+3\lambda}{2}
\end{smallmatrix};1\right]\right|_{\nu=\lambda} \\&\qquad +3\,_{7}F_{6}\left[\begin{smallmatrix}1,\lambda,\lambda,\lambda,\lambda,\frac{1+\lambda}{2},\frac{1+\lambda}{2}\\
1+\lambda,1+\lambda,1+\lambda,1+\lambda,\frac{1+3\lambda}{2},\frac{1+3\lambda}{2}
\end{smallmatrix};1\right].
\end{aligned}
\end{equation}
In summary, the Wilson coefficients are
\be \hspace{-1mm}
\begin{aligned}
g_0 =& 2\,{}_4 F_3 \left[\begin{smallmatrix}1,\lambda,\frac{1+\lambda}{2},\frac{1+\lambda}{2} \\ 1+\lambda,\frac{1+3\lambda}{2},\frac{1+3\lambda}{2} \end{smallmatrix};1\right]\\
g_2 =& \,{}_6 F_5 \left[\begin{smallmatrix}1,\lambda,\lambda,\lambda,\frac{1+\lambda}{2},\frac{1+\lambda}{2} \\ 1+\lambda,1+\lambda,1+\lambda,\frac{1+3\lambda}{2},\frac{1+3\lambda}{2} \end{smallmatrix};1\right] \\
g_3 =& -4\left.\tfrac{{\rm d}}{{\rm d} \log \nu}{}_{6}F_{5}\left[\begin{smallmatrix}
1,\lambda,\lambda,\lambda,\frac{1+\nu}{2},\frac{1+\nu}{2}\\
1+\lambda,1+\lambda,1+\lambda,\frac{1+3\lambda}{2},\frac{1+3\lambda}{2}
\end{smallmatrix};1\right]\right|_{\nu=\lambda} \\ & +3 \,{}_7 F_6 \left[\begin{smallmatrix}1,\lambda,\lambda,\lambda,\lambda,\frac{1+\lambda}{2},\frac{1+\lambda}{2} \\ 1+\lambda,1+\lambda,1+\lambda,1+\lambda,\frac{1+3\lambda}{2},\frac{1+3\lambda}{2} \end{smallmatrix};1\right]\\
g_4 =& \frac{1}{2}\,{}_8 F_7 \left[\begin{smallmatrix}1,\lambda,\lambda,\lambda,\lambda,\lambda,\frac{1+\lambda}{2},\frac{1+\lambda}{2} \\ 1+\lambda,1+\lambda,1+\lambda,1+\lambda,1+\lambda,\frac{1+3\lambda}{2},\frac{1+3\lambda}{2} \end{smallmatrix};1\right].
\end{aligned} 
\ee

The Regge limit of $M$ at large $s$ and fixed $t$ is
\be
M \sim s^{2J(s,t)} + \frac{1}{s^2}\left[-\frac{\rho^2}{t} + \frac{(3\lambda-1)^2}{4\lambda(1-\lambda+\lambda t)}\right]+\cdots,
\label{eq:MRegge}
\ee
where $J(s,t) = \lambda(t-1)+\cdots$.
The form of Eq.~\eqref{eq:MRegge} can be deduced semianalytically by computing the Regge limit at various choices of $t=\xi(k)$, where by construction the amplitude becomes a rational polynomial in $s$; the expression in Eq.~\eqref{eq:MRegge} can then be numerically verified to high precision for general values of $t$.
When $t>\tfrac{\lambda-1}{\lambda}$, the amplitude is dominated by the $s^{2J(s,t)}$ term, while for $t<\tfrac{\lambda-1}{\lambda}$, $M\sim 1/s^2$, with 
the spurious pole at $t=\tfrac{\lambda-1}{\lambda}$ indicating the transition.
In the hard scattering limit, i.e., $|s|,|t|\rightarrow\infty$ with $t/s$ fixed,
\be
M  \sim e^{2\lambda B(s,t)} + \frac{1}{stu}\left[\rho^2-\left(\frac{3\lambda-1}{2\lambda}\right)^2\right]+\cdots,\label{eq:Mhard}
\ee
where $B(s,t) = -s\log s-t\log t-u\log u+\cdots$, and where the second or first term in Eq.~\eqref{eq:Mhard} dominates when the scattering angle defined by $\cos\theta=1+\tfrac{2t}{s}$ is physical---i.e., $\cos\theta \in [-1,1]$---or unphysical, respectively. 
As for the Regge limit, we found the hard scattering limit in Eq.~\eqref{eq:Mhard} semianalytically, explicitly computing the hard scattering limit using various rational values of $\lambda$ and $t/s$ at which the hypergeometric function simplifies, deducing the pattern, and then numerically verifying Eq.~\eqref{eq:Mhard} for general $t/s$ and $\lambda$.

To describe graviton scattering, we  dress $M$ with an additional factor of ${\cal R}^4$ that accounts for the external polarizations~\footnote{The graviton amplitude in general relativity is $8\pi G\,{\cal R}^4/stu$, where 
the expression for ${\cal R}^4$  
can be found in, e.g., Ref.~\cite{Arkani-Hamed:2022gsa}.}.  For generic $\rho$ and $\lambda$, the Regge limit of $M$ defined in Eq.~\eqref{eq:MRegge} scales as $s^2$  for $t<\tfrac{\lambda-1}{\lambda}$.   This saturates the causality bounds on gravitational Regge behavior~\cite{Arkani-Hamed:2020blm,CEMZ,Haring:2022cyf} and scales no better than graviton scattering in general relativity.  Hence, these variations of VS are not bona fide ultraviolet completions of gravity, though for the special choice of $\rho=\tfrac{3\lambda-1}{2\lambda}$ the hard scattering limit is improved over pure general relativity, scaling as $s^0$ rather than $s^1$.

As another consistency check, let us study our amplitudes in the soft limit in which the momentum $q$ of a single external leg is taken to zero.  Since ${\cal R}^4 \sim O(q^2)$, the leading $1/stu \sim 1/q^3$ term in $M$ correctly yields the leading graviton soft theorem~\cite{Weinberg:1965nx,Weinberg:1964ew}.
Meanwhile, there is no term at $O(1/q^2)$ in $M$, so the subleading graviton soft theorem holds as well, as required by unitarity, locality, and CPT invariance~\cite{Elvang:2016qvq}.  While the subsubleading graviton soft theorem is not universal, we find that the $O(1/q)$ term of $M$ is also zero, so this behavior is unchanged from general relativitiy.

Of course, these amplitudes can also be interpreted as describing the scattering of indistinguishable scalars, independent of gravity.  In this case there are no polarization-dependent prefactors, but we should fix $\rho\,{=}\,0$ so that there are no nonlocal poles.  The resulting scalar amplitudes appear to be perfectly sensible, in line with the kinds of amplitudes constrained in Ref.~\cite{Caron-Huot:2020cmc}.
In this case, the $O(1/s^2)$ behavior of the amplitude in the Regge limit is an improvement over the behavior of massive scalar exchange, which is remarkable given the presence of an infinite tower of massive spinning states.

\medskip

\begin{figure*}
\begin{minipage}[t]{0.29\textwidth}
\includegraphics[width=\textwidth]{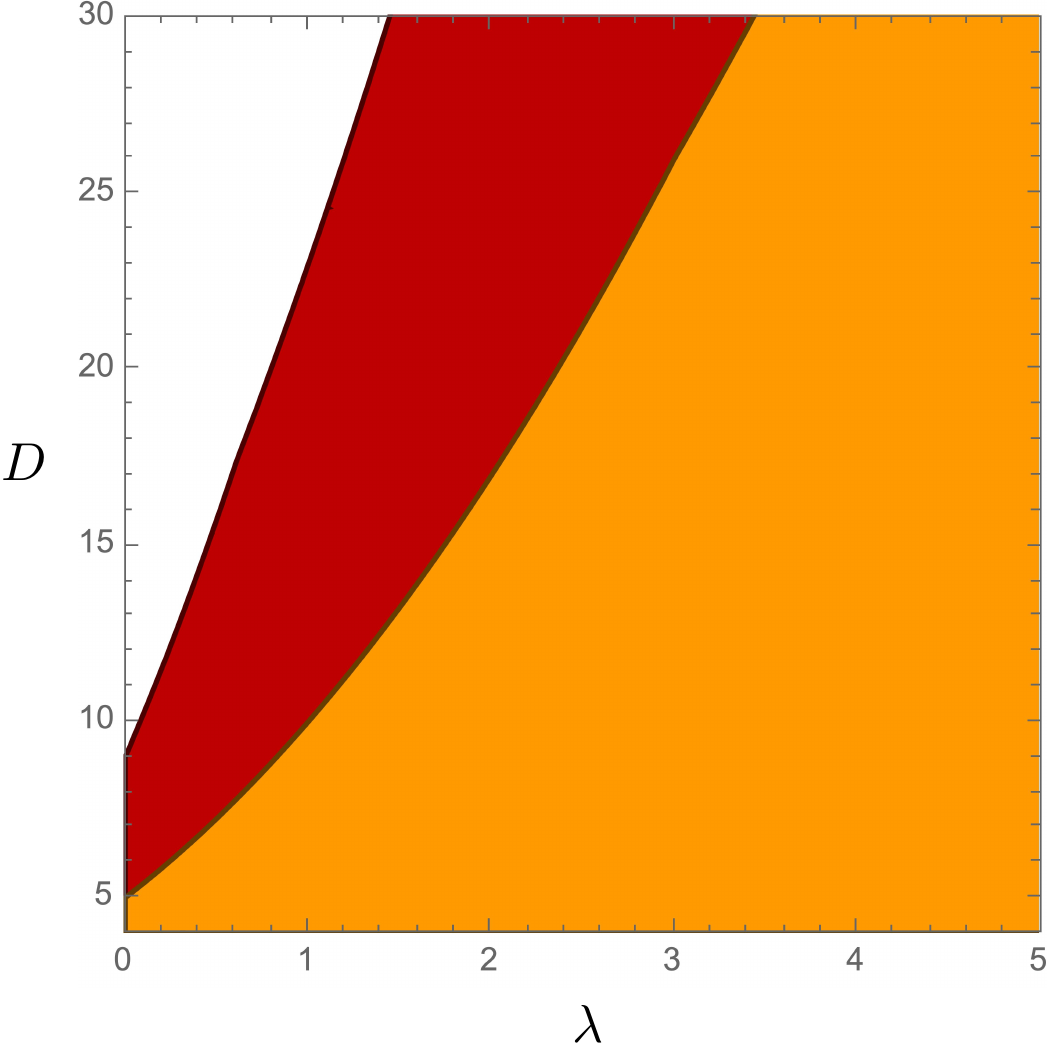}
\vspace{-7mm}
\caption{Values of the parameter $\lambda$ and spacetime dimension $D$ consistent with positivity, with nonnegativity of partial waves verified  through level $n\,{=}\,30$.
Orange: $M$ in Eq.~\eqref{eq:M} and $A$ in Eq.~\eqref{eq:A} are both positive.
Red: additional region where $M$ alone is positive.}\label{fig:positivity}
\end{minipage}%
\hfill
\begin{minipage}[t]{0.67\textwidth}
\includegraphics[width=\textwidth]{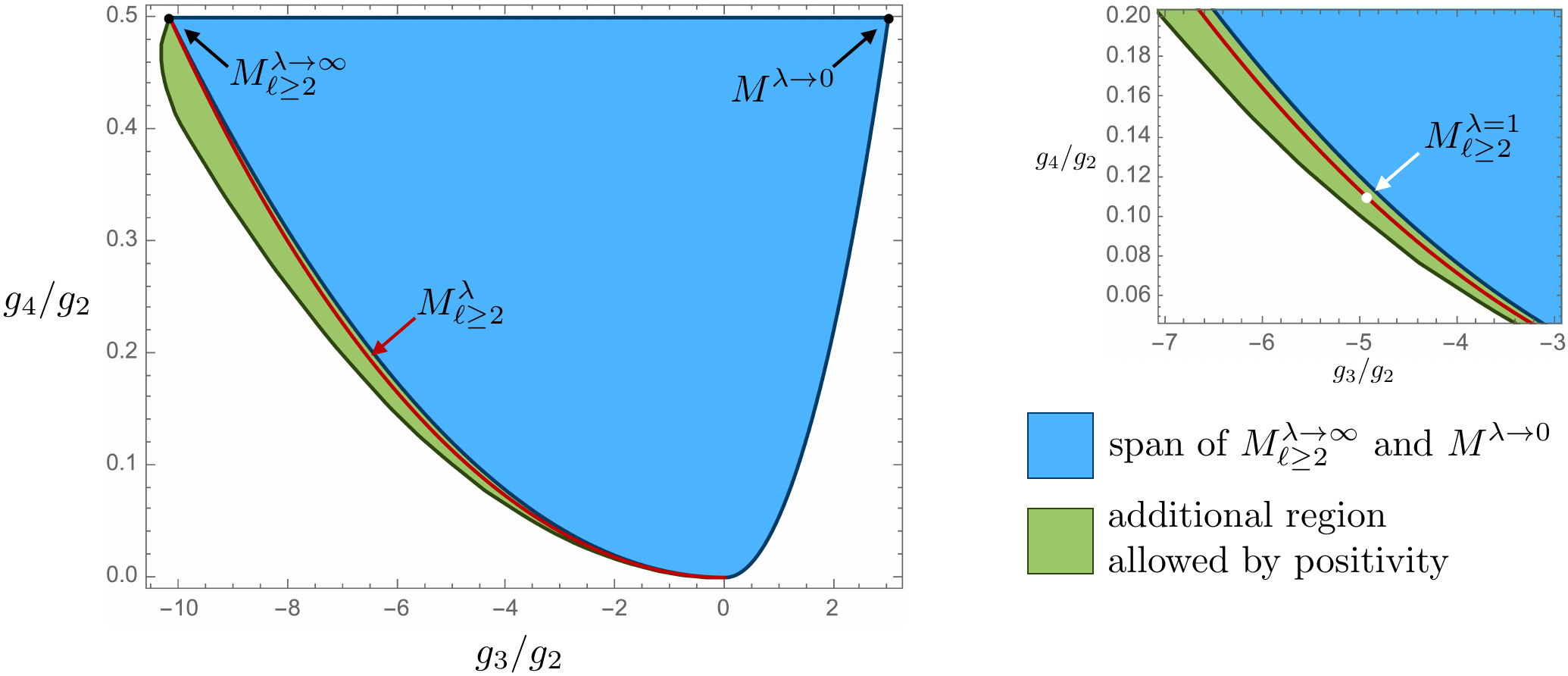}
\vspace{-7mm}
\caption{Extremal EFT couplings (black dots) are generated  by the $stu$-pole amplitude, with scalar partial waves subtracted, and by the scalar exchange amplitude.
The convex hull (blue) of these extremal amplitudes is generated by rescaling the mass. 
The full space of EFT couplings consistent with unitarity~\cite{Caron-Huot:2020cmc} extends to the additional green region.
Both extremal corners are generated by our amplitude $M$ in Eq.~\eqref{eq:M}, for $\lambda\,{\rightarrow}\,\infty$ in Eq.~\eqref{eq:stupole}, with $\ell\,{=}\,0$ states removed, and $\lambda\,{\rightarrow}\,0$ in Eq.~\eqref{eq:scalar}, respectively.
At arbitrary $\lambda$, subtracting scalar partial waves from $M$ generates the red curve for different parameter values, exiting the blue region into the green sliver. 
}\label{fig:EFT}
\end{minipage}
\end{figure*}

\mysubsec{Positivity}As a necessary condition for unitarity, in particular the absence of ghosts, we must be able to expand the residues in partial waves, $R(n,t) = \sum_{\ell=0}^{2n-2} a_{n,\ell} G_\ell^{(D)}(\cos\theta)$,
with all nonnegative coefficients $a_{n,\ell}$, where the $G_\ell^{(D)}$ are Gegenbauer polynomials describing spherical harmonics in $D$ spacetime dimensions.
For the residues in Eq.~\eqref{eq:Resfinal}, with the help of identities in Ref.~\cite{Cheung:2022mkw}, we can compute the partial wave coefficients~\footnote{We find the partial wave coefficients for $R(n,t)$, describing the amplitude $M$, to be given by\vspace{-0.5mm}
\begin{equation*}\hspace{7.5mm}
\begin{aligned}
a_{n,\ell}&=\tfrac{\left(1+\frac{2\ell}{D-3}\right)\Gamma\left(\frac{D-1}{2}\right)}{\left(\frac{1+3\lambda}{2}\right)_{n-1}^{2}}\SigmaScaled_{i=0}^{n-1}\SigmaScaled_{j=\ell}^{2n-2-i}\SigmaScaled_{k=0}^{\lfloor(j-\ell)/2\rfloor}a_{n,\ell}^{i,j,k}\\
a_{n,\ell}^{i,j,k} &=
\tfrac{(-1)^{i}(n-i)_i^2 S_{1}(2n-2-i,j)}{i!k!(j-\ell-2k)!}\tfrac{j!(2-n)^{j-\ell-2k}(n+\lambda-1)^{\ell+2k}}{2^{j+\ell+2k}\Gamma\left(\frac{D-1}{2}+\ell+k\right)},
\end{aligned}
\end{equation*}
where $S_1$ are unsigned Stirling numbers of the first kind. For $n\,{=}\,2$, $(2-n)^{j-\ell-2k}$ should be replaced with $\delta_{2k+\ell,2}$.}.
As for VS, $a_{n,\ell} = 0$ for $\ell$ odd.
In $D=4$ dimensions, we find positivity for all $\lambda \geq 0$.
When $D\geq 9$, positivity bounds $\lambda$ from below, as in Fig.~\ref{fig:positivity}.

\bigskip

\mysec{Extremal Limits}In addition to representing a generalized family of amplitudes containing the closed string, Eq.~\eqref{eq:M} exhibits extraordinary behavior in extreme parameter limits.
In particular, when $\lambda\rightarrow 0$ the higher-spin states decouple, leaving us with gravity plus massive scalar exchange,
\be
M(s,t)\stackrel{\lambda\rightarrow 0}{\longrightarrow}\frac{\rho^2}{stu}+\frac{1}{1-s}+\frac{1}{1-t}+\frac{1}{1-u}\label{eq:scalar}.
\ee
While Eq.~\eqref{eq:scalar} is clearly not dual resonant on its own, since it does not vanish in the Regge limit, $M$ is still dual resonant for any $\lambda>0$; this is possible because the higher-spin poles that ensure dual resonance are pushed parametrically into the ultraviolet when $\lambda\rightarrow 0$.

For $\lambda\rightarrow\infty$, the amplitudes takes the striking form, 
\be
M(s,t)\stackrel{\lambda\rightarrow\infty}{\longrightarrow}  \frac{\rho^2}{stu} + \frac{9}{4(1-s)(1-t)(1-u)}.\label{eq:stupole}
\ee
Here, the entire tower of string excitations has collapsed to a single resonance at $\mu=1$, whose residue is nonlocal.

These limits are remarkable, as they constitute the kinks in the parameter space of allowed EFT couplings determined in Ref.~\cite{Caron-Huot:2020cmc}. In the case of Eq.~\eqref{eq:stupole}---the so-called ``$stu$-pole'' amplitude---the scalar partial waves must be subtracted to realize the kink~\cite{Caron-Huot:2020cmc}. In particular, in the scalar-subtracted case, the $\lambda$ parameter interpolates between the ``higher-spin'' kink and the origin as $\lambda$ varies from $\infty$ to $0$, as shown in Fig.~\ref{fig:EFT}.  Even more notably, the trajectory in coefficient space exits the region analytically ruled in by the $stu$-pole and scalar exchange theories alone.  Perhaps an approach akin to the $\lambda$ deformation studied here could produce the boundary found with semidefinite programming, which has thus far eluded analytic understanding in the literature.

The ${}_5 F_4$ hypergeometric function in $M$ is performing a nontrivial mathematical miracle here: dropping the ${}_5 F_4$ function in Eq.~\eqref{eq:M} by setting it to unity---in effect, considering the VS amplitude with shifted and rescaled gamma function arguments---leads to an ill-defined $\lambda\rightarrow\infty$ limit.
The delicate structure of $M$, a consequence of level truncation, appears key to recovering Eq.~\eqref{eq:stupole}.

\bigskip

\mysec{Planar Analogue}Let us try to construct a planar version of our amplitude in Eq.~\eqref{eq:M}.
That is, we might ask: does there exist a deformation of the Veneziano amplitude, rather than VS, with the spectrum given in Eq.~\eqref{eq:sols}?
Motivated by string theory and the double copy, let us posit polynomial residues $\hat R(n,t)$ of degree $n-1$ for the planar amplitude, given by the square root of that of our gravity amplitude in Eq.~\eqref{eq:Resfinal}, so that $\hat R(0,t) = \rho/t$ and
\be
\hat R(n,t) = \frac{\left(\frac{1+\lambda}{2}+\lambda t\right)_{n-1}}{\left(\frac{1+3\lambda}{2}\right)_{n-1}} = \sqrt{R(n,t)}.\label{eq:resplanar}
\ee

\medskip
\mysubsec{Amplitudes and Extremal Limits}Starting with the dual resonant sum,
\be 
A(s,t) = \sum_{n=0}^\infty \frac{\hat R(n,t)}{\mu(n) - s} 
\ee
we find via hypergeometric identities that the amplitude can be expressed in manifestly crossing symmetric form,
\be 
\hspace{-2mm}\begin{aligned}
&A(s,t) =-\frac{\rho}{st}\\&+\frac{\lambda\Gamma(\lambda-\lambda s)\Gamma(\lambda -\lambda t)}{\Gamma(2\lambda -\lambda s-\lambda t)}\,_{3}F_{2}\left[\begin{smallmatrix}
\lambda(1{-}s),\lambda(1{-}t),\frac{3\lambda{-}1}{2}\\
\lambda(2{-}s{-}t),\frac{3\lambda{+}1}{2}
\end{smallmatrix};1\right].\label{eq:A}
\end{aligned}\hspace{-2mm}
\ee
Remarkably---and unlike the deformation of VS we found in Eq.~\eqref{eq:M}---we recognize this object.
It is a variation on the hypergeometric amplitude discovered in Ref.~\cite{Cheung:2023adk}, with the Mandelstam variables shifted by $1$ and then rescaled by $\lambda$, the massless pole added back in, and the hypergeometric $r$ parameter set to $\tfrac{3\lambda-1}{2}$.
Like $M$ in Eq.~\eqref{eq:M}, $A$ reduces to either scalar exchange or an infinite tower of spins in extreme limits of~$\lambda$,
\eq{
A &\stackrel{\lambda\rightarrow 0}{\longrightarrow} -\frac{\rho}{st} + \frac{1}{1-s} + \frac{1}{1-t}\\
A&\stackrel{\lambda\rightarrow\infty}{\longrightarrow} -\frac{\rho}{st} + \frac{3}{2(1-s)(1-t)},
}{eq:planarscalarpole}
while for $\lambda=\rho=1$, we obtain the superstring version of the Veneziano amplitude,
\be
A\stackrel{\lambda=\rho=1}{\longrightarrow}  -\frac{\Gamma(-s)\Gamma(-t)}{\Gamma(1-s-t)}.\label{eq:V}
\ee

\medskip

\mysubsec{Positivity and Kinematic Limits}At low energies, $A$ has the EFT expansion,
\be
A = -\frac{\rho}{st} + {}_3 F_2 \left[\begin{smallmatrix}1,\lambda,\frac{1+\lambda}{2}\\ 1+\lambda,\frac{1+3\lambda}{2} \end{smallmatrix};1\right] + \cdots.
\ee
In superstring theory, the Veneziano amplitude in Eq.~\eqref{eq:V} is dressed with a kinematic prefactor ${\cal F}^4$ containing polarization data.
Taking one leg soft, ${\cal F}^4 \sim O(q)$, while $A$ is dominated by $1/st \sim 1/q^2$, so the leading gluon soft theorem is unchanged from Yang-Mills theory~\cite{Weinberg:1965nx,Weinberg:1964ew}.
The $O(1/q)$ part of $A$ is zero, so the subleading soft theorem---which in any case is not universal~\cite{Elvang:2016qvq}---is also unchanged.  The  first deviation in the soft behavior appears at $O(q^0)$ in $A$, which is subsubleading in the soft limit.

At high energies, $A(s,t)$ behaves like 
\be 
\begin{aligned}
A_{\rm Regge} &\sim s^{J(s,t)} + \frac{1}{s}\left[-\frac{\rho}{t} + \frac{3\lambda-1}{2(1-\lambda+\lambda t)}\right]+\cdots \\
A_{\rm hard} &\sim e^{\lambda B(s,t)} + \frac{1}{st}\left(-\rho+\frac{3\lambda-1}{2\lambda}\right)+\cdots
\end{aligned}
\ee
where $J(s,t)$ and $B(s,t)$ are as in Eqs.~\eqref{eq:MRegge} and \eqref{eq:Mhard}.
As in Eq.~\eqref{eq:Msatellite}, we can write $A$ as an infinite satellite sum over Veneziano amplitudes with rescaled momenta,
\be
A = -\frac{\rho}{st} {+} \sum_{k=0}^\infty \frac{\lambda(3\lambda{-}1)\Gamma(k{+}\lambda {-}\lambda s)\Gamma(k{+}\lambda {-} \lambda t)}{(3\lambda{-}1{+}2k)k!\Gamma(k{+}2\lambda{-}\lambda s{-}\lambda t)}.
\ee
Expanding the residue $\hat R(n,t)$ in partial waves as $\sum_{\ell=0}^{n-1} \hat a_{n,\ell} G_\ell^{(D)}(\cos\theta)$~\footnote{We find the partial wave coefficients for $\hat R(n,t)$, describing the amplitude $A$, to be given by\vspace{-0.5mm}
\begin{equation*}\hspace{7.5mm}
\begin{aligned}
\hat a_{n,\ell}  &=\tfrac{\left(1+\frac{2\ell}{D-3}\right)\Gamma\left(\frac{D-1}{2}\right)}{\left(\frac{1+3\lambda}{2}\right)_{n-1}}\SigmaScaled_{j=\ell}^{n-1}\SigmaScaled_{k=0}^{\lfloor(j-\ell)/2\rfloor} \hat a_{n,\ell}^{j,k} \\
\hat a_{n,\ell}^{j,k}&=\tfrac{S_{1}(n-1,j)}{k!(j-\ell-2k)!}\tfrac{j!(2-n)^{j-\ell-2k}(n+\lambda -1)^{\ell+2k}}{2^{j+\ell+2k}\Gamma\left(\frac{D-1}{2}+\ell+k\right)}.
\end{aligned}\vspace{-1.5mm}
\end{equation*}
For $n=2$, $(2-n)^{j-\ell-2k}$ should be replaced with $\delta_{\ell,1}$.},
we find positivity in $D\,{=}\,4$ dimensions for all $\lambda \geq 0$. When $D\geq 5$, there is a minimum value of $\lambda$ for which $\hat a_{n,\ell} \geq 0$; see Fig.~\ref{fig:positivity}.

Finally it is worth remarking that as required by Eq.~\eqref{eq:planarscalarpole}, the crossing symmetric sum $A(s,t)+A(t,u)+A(u,s)$ has the aforementioned extremal EFTs as its extreme limits in $\lambda$. Indeed, expanding at low energies, we obtain a curve that enters the green region of EFT coupling space in Fig.~\ref{fig:EFT}, very close to but distinct from that of the VS deformation $M$ for different values of $\lambda$. 

\medskip

\mysec{Discussion}In this paper, we have applied level truncation to the closed string.
In Ref.~\cite{Cheung:2024uhn}, we showed that the Veneziano amplitude of the open string can be uniquely bootstrapped---including its spectrum---by demanding level truncation and {\it superpolynomial softness}~\cite{Haring:2023zwu}, the requirement that for any positive integer $k$ there exists some range of $t$ for which the amplitude falls off faster than $s^{-k}$ in the Regge limit.
In the present paper, we bootstrapped the unique two-parameter family of permutation invariant, dual resonant amplitudes obeying level truncation with double zeros as in Eq.~\eqref{eq:Ransatz}, with the result given in Eq.~\eqref{eq:M}.  As before, the spectrum is an output of the calculation.
From Eq.~\eqref{eq:MRegge}, we see that the only choice of parameters for which superpolynomial softness holds for a range of $t$ is $\lambda=\rho=1$.
That is, the unique amplitude obeying level truncation of $M$ and $\partial_t M$, along with superpolynomial softness, is the VS amplitude.
This is the first analytic bootstrap construction that uniquely produces the graviton scattering amplitude of string theory from physical constraints~\footnote{Analytic bootstraps have appeared that assume a product ansatz.  However, this is not a physical property of scattering, but rather just the form of the function.  In particular, a physical measurement cannot determine whether an amplitude is expressible in product form.}.

The fact that the amplitude $M$ in Eq.~\eqref{eq:M} reproduces the extremal EFT amplitudes of Ref.~\cite{Caron-Huot:2020cmc} for extreme values of $\lambda$ makes it a compelling object for future study.
Further, the amplitude $A$ in Eq.~\eqref{eq:A} may also represent an extremization from an EFT perspective, even at finite $\lambda$.
In recent work~\cite{Berman:2024wyt}, a string-inspired bootstrap for the EFT coefficients of planar amplitudes was investigated, and it was found that a spectrum precisely of the form in Eq.~\eqref{eq:sols}---with massless external states and a linear spectrum with arbitrary slope---numerically extremizes the coefficients, which was conjectured to correspond to some unknown ``corner theory'' amplitude.
We also note that the the planar versions of the extremal amplitudes, which our $A$ produces for extreme $\lambda$ in Eq.~\eqref{eq:planarscalarpole}, occupy kinks in the allowed EFT parameter space~\cite{Haring:2023zwu}.
We leave an analysis of the relations between our amplitudes and positivity to future work.
Other compelling avenues for further investigation include generalizing our results to higher-point scattering and investigating amplitudes with modified versions of level truncation such as those in Ref.~\cite{Bjerrum-Bohr:2024wyw}. 
The question of the physical justification for level truncation itself, which has proven so powerful in uniquely bootstrapping the string, remains open.

\bigskip
\vspace{-1mm}

\noindent {\it Acknowledgments:} 
We thank Andrea Guerrieri for useful discussions. 
 C.C. and A.H. are supported by the Department of Energy (Grant No.~DE-SC0011632) and by the Walter Burke Institute for Theoretical Physics.
G.N.R. is supported by the James Arthur Postdoctoral Fellowship at New York University.

\bibliographystyle{utphys-modified}
\bibliography{VS}

\providecommand{\href}[2]{#2}\begingroup\raggedright\begin{thebibliography}{10}

\bibitem{Veneziano}
G.~Veneziano, ``{Construction of a crossing-symmetric, Regge-behaved amplitude
  for linearly rising trajectories},''
  \href{http://dx.doi.org/10.1007/BF02824451}{{\em Nuovo Cim. A} {\bfseries 57}
  (1968) 190}.

\bibitem{Coon:1969yw}
D.~D. Coon, ``{Uniqueness of the Veneziano representation},''
  \href{http://dx.doi.org/10.1016/0370-2693(69)90106-3}{{\em Phys. Lett. B}
  {\bfseries 29} (1969) 669}.

\bibitem{Gross:1969db}
D.~J. Gross, ``{Factorization and the generalized Veneziano model with
  satellites},'' \href{http://dx.doi.org/10.1016/0550-3213(69)90248-X}{{\em
  Nucl. Phys. B} {\bfseries 13} (1969) 467}.

\bibitem{Mandelstam:1968czc}
S.~Mandelstam, ``{Veneziano Formula with Trajectories Spaced by Two Units},''
  \href{http://dx.doi.org/10.1103/PhysRevLett.21.1724}{{\em Phys. Rev. Lett.}
  {\bfseries 21} (1968) 1724}.

\bibitem{Cheung:2022mkw}
C.~Cheung and G.~N. Remmen, ``{Veneziano variations: how unique are string
  amplitudes?},'' \href{http://dx.doi.org/10.1007/JHEP01(2023)122}{{\em JHEP}
  {\bfseries 01} (2023) 122}, \href{http://arxiv.org/abs/2210.12163}{{\ttfamily
  arXiv:2210.12163 [hep-th]}}.

\bibitem{Cheung:2023adk}
C.~Cheung and G.~N. Remmen, ``{Stringy dynamics from an amplitudes
  bootstrap},'' \href{http://dx.doi.org/10.1103/PhysRevD.108.026011}{{\em Phys.
  Rev. D} {\bfseries 108} (2023) 026011},
  \href{http://arxiv.org/abs/2302.12263}{{\ttfamily arXiv:2302.12263
  [hep-th]}}.

\bibitem{Cheung:2023uwn}
C.~Cheung and G.~N. Remmen, ``{Bespoke dual resonance},''
  \href{http://dx.doi.org/10.1103/PhysRevD.108.086009}{{\em Phys. Rev. D}
  {\bfseries 108} (2023) 086009},
  \href{http://arxiv.org/abs/2308.03833}{{\ttfamily arXiv:2308.03833
  [hep-th]}}.

\bibitem{Haring:2023zwu}
K.~H\"aring and A.~Zhiboedov, ``{The Stringy S-matrix Bootstrap: Maximal Spin
  and Superpolynomial Softness},''
  \href{http://arxiv.org/abs/2311.13631}{{\ttfamily arXiv:2311.13631
  [hep-th]}}.

\bibitem{Bhardwaj:2024klc}
R.~Bhardwaj, M.~Spradlin, A.~Volovich, and H.-C. Weng, ``{On Unitarity of
  Bespoke Amplitudes},'' \href{http://arxiv.org/abs/2406.04410}{{\ttfamily
  arXiv:2406.04410 [hep-th]}}.

\bibitem{Maldacena:2022ckr}
J.~Maldacena and G.~N. Remmen, ``{Accumulation-point amplitudes in string
  theory},'' \href{http://dx.doi.org/10.1007/JHEP08(2022)152}{{\em JHEP}
  {\bfseries 08} (2022) 152}, \href{http://arxiv.org/abs/2207.06426}{{\ttfamily
  arXiv:2207.06426 [hep-th]}}.

\bibitem{Bjerrum-Bohr:2024wyw}
N.~E.~J. Bjerrum-Bohr and C.~B. Jepsen, ``{Scattering on the Worldvolume:
  Amplitude Relations in Brower-Goddard String Models},''
  \href{http://arxiv.org/abs/2406.10176}{{\ttfamily arXiv:2406.10176
  [hep-th]}}.

\bibitem{Bhardwaj:2023eus}
R.~Bhardwaj and S.~De, ``{Dual resonant amplitudes from Drinfel'd twists},''
  \href{http://arxiv.org/abs/2309.07214}{{\ttfamily arXiv:2309.07214
  [hep-th]}}.

\bibitem{Eckner:2024ggx}
C.~Eckner, F.~Figueroa, and P.~Tourkine, ``{The Regge bootstrap, from linear to
  non-linear trajectories},'' \href{http://arxiv.org/abs/2401.08736}{{\ttfamily
  arXiv:2401.08736 [hep-th]}}.

\bibitem{Eckner:2024pqt}
C.~Eckner, F.~Figueroa, and P.~Tourkine, ``{On the number of Regge trajectories
  for dual amplitudes},'' \href{http://arxiv.org/abs/2405.21057}{{\ttfamily
  arXiv:2405.21057 [hep-th]}}.

\bibitem{Geiser:2022icl}
N.~Geiser and L.~W. Lindwasser, ``{Properties of infinite product amplitudes:
  Veneziano, Virasoro, and Coon},''
  \href{http://dx.doi.org/10.1007/JHEP12(2022)112}{{\em JHEP} {\bfseries 12}
  (2022) 112}, \href{http://arxiv.org/abs/2207.08855}{{\ttfamily
  arXiv:2207.08855 [hep-th]}}.

\bibitem{Geiser:2022exp}
N.~Geiser and L.~W. Lindwasser, ``{Generalized Veneziano and Virasoro
  amplitudes},'' \href{http://dx.doi.org/10.1007/JHEP04(2023)031}{{\em JHEP}
  {\bfseries 04} (2023) 031}, \href{http://arxiv.org/abs/2210.14920}{{\ttfamily
  arXiv:2210.14920 [hep-th]}}.

\bibitem{Figueroa:2022onw}
F.~Figueroa and P.~Tourkine, ``{Unitarity and Low Energy Expansion of the Coon
  Amplitude},'' \href{http://dx.doi.org/10.1103/PhysRevLett.129.121602}{{\em
  Phys. Rev. Lett.} {\bfseries 129} (2022) 121602},
  \href{http://arxiv.org/abs/2201.12331}{{\ttfamily arXiv:2201.12331
  [hep-th]}}.

\bibitem{Chakravarty:2022vrp}
J.~Chakravarty, P.~Maity, and A.~Mishra, ``{On the positivity of Coon amplitude
  in $D = 4$},'' \href{http://dx.doi.org/10.1007/JHEP10(2022)043}{{\em JHEP}
  {\bfseries 10} (2022) 043}, \href{http://arxiv.org/abs/2208.02735}{{\ttfamily
  arXiv:2208.02735 [hep-th]}}.

\bibitem{Geiser:2023qqq}
N.~Geiser, ``{The Baker-Coon-Romans $N$-point amplitude and an exact field
  theory limit of the Coon amplitude},''
  \href{http://arxiv.org/abs/2311.04130}{{\ttfamily arXiv:2311.04130
  [hep-th]}}.

\bibitem{Jepsen:2023sia}
C.~B. Jepsen, ``{Cutting the Coon amplitude},''
  \href{http://dx.doi.org/10.1007/JHEP06(2023)114}{{\em JHEP} {\bfseries 06}
  (2023) 114}, \href{http://arxiv.org/abs/2303.02149}{{\ttfamily
  arXiv:2303.02149 [hep-th]}}.

\bibitem{Bhardwaj:2022lbz}
R.~Bhardwaj, S.~De, M.~Spradlin, and A.~Volovich, ``{On unitarity of the Coon
  amplitude},'' \href{http://dx.doi.org/10.1007/JHEP08(2023)082}{{\em JHEP}
  {\bfseries 08} (2023) 082}, \href{http://arxiv.org/abs/2212.00764}{{\ttfamily
  arXiv:2212.00764 [hep-th]}}.

\bibitem{Virasoro}
M.~A. Virasoro, ``{Alternative Constructions of Crossing-Symmetric Amplitudes
  with Regge Behavior},''
  \href{http://dx.doi.org/10.1103/PhysRev.177.2309}{{\em Phys. Rev.} {\bfseries
  177} (1969) 2309}.

\bibitem{Shapiro}
J.~A. Shapiro, ``{Electrostatic analogue for the Virasoro model},''
  \href{http://dx.doi.org/10.1016/0370-2693(70)90255-8}{{\em Phys. Lett. B}
  {\bfseries 33} (1970) 361}.

\bibitem{Note1}
Here we have stripped off the prefactor involving graviton polarizations and
  have chosen convenient units for the string tension.

\bibitem{Chowdhury:2019kaq}
S.~D. Chowdhury, A.~Gadde, T.~Gopalka, I.~Halder, L.~Janagal, and S.~Minwalla,
  ``{Classifying and constraining local four photon and four graviton
  S-matrices},'' \href{http://dx.doi.org/10.1007/JHEP02(2020)114}{{\em JHEP}
  {\bfseries 02} (2020) 114}, \href{http://arxiv.org/abs/1910.14392}{{\ttfamily
  arXiv:1910.14392 [hep-th]}}.

\bibitem{Guerrieri:2021ivu}
A.~Guerrieri, J.~Penedones, and P.~Vieira, ``{Where Is String Theory in the
  Space of Scattering Amplitudes?},''
  \href{http://dx.doi.org/10.1103/PhysRevLett.127.081601}{{\em Phys. Rev.
  Lett.} {\bfseries 127} (2021) 081601},
  \href{http://arxiv.org/abs/2102.02847}{{\ttfamily arXiv:2102.02847
  [hep-th]}}.

\bibitem{Caron-Huot:2021rmr}
S.~Caron-Huot, D.~Mazac, L.~Rastelli, and D.~Simmons-Duffin, ``{Sharp
  boundaries for the swampland},''
  \href{http://dx.doi.org/10.1007/JHEP07(2021)110}{{\em JHEP} {\bfseries 07}
  (2021) 110}, \href{http://arxiv.org/abs/2102.08951}{{\ttfamily
  arXiv:2102.08951 [hep-th]}}.

\bibitem{Bern:2021ppb}
Z.~Bern, D.~Kosmopoulos, and A.~Zhiboedov, ``{Gravitational effective field
  theory islands, low-spin dominance, and the four-graviton amplitude},''
  \href{http://dx.doi.org/10.1088/1751-8121/ac0e51}{{\em J. Phys. A} {\bfseries
  54} (2021) 344002}, \href{http://arxiv.org/abs/2103.12728}{{\ttfamily
  arXiv:2103.12728 [hep-th]}}.

\bibitem{Huang:2022mdb}
Y.-t. Huang and G.~N. Remmen, ``{UV-complete gravity amplitudes and the triple
  product},'' \href{http://dx.doi.org/10.1103/PhysRevD.106.L021902}{{\em Phys.
  Rev. D} {\bfseries 106} (2022) L021902},
  \href{http://arxiv.org/abs/2203.00696}{{\ttfamily arXiv:2203.00696
  [hep-th]}}.

\bibitem{Arkani-Hamed:2020blm}
N.~Arkani-Hamed, T.-C. Huang, and Y.-t. Huang, ``{The EFT-hedron},''
  \href{http://dx.doi.org/10.1007/JHEP05(2021)259}{{\em JHEP} {\bfseries 05}
  (2021) 259}, \href{http://arxiv.org/abs/2012.15849}{{\ttfamily
  arXiv:2012.15849 [hep-th]}}.

\bibitem{Albert:2024yap}
J.~Albert, W.~Knop, and L.~Rastelli, ``{Where is tree-level string theory?},''
  \href{http://arxiv.org/abs/2406.12959}{{\ttfamily arXiv:2406.12959
  [hep-th]}}.

\bibitem{Cheung:2024uhn}
C.~Cheung, A.~Hillman, and G.~N. Remmen, ``{A Bootstrap Principle for the
  Spectrum and Scattering of Strings},''
  \href{http://arxiv.org/abs/2406.02665}{{\ttfamily arXiv:2406.02665
  [hep-th]}}.

\bibitem{Note2}
In contrast, Ref.~\cite {Caron-Huot:2016icg} showed that the open string
  amplitude was {\protect \it asymptotically} unique under broad technical
  assumptions that have recently been shown to be violated in certain
  cases~\cite {Haring:2023zwu}. Alternatively, adding the strong assumption of
  string monodromy allows for the bootstrapping of Veneziano from dispersion
  relations~\cite {Berman:2023jys,Chiang:2023quf}. Historically, Coon~\cite
  {Coon:1969yw} bootstrapped the spectrum of his $q$-deformation of the
  Veneziano amplitude by requiring infinite product form of the bootstrap.

\bibitem{Note3}
Straightforward $\hbar $ counting is sufficient to show that any perturbative
  ultraviolet completion of gravity must be implemented by tree-level
  exchanges~\cite {Cheung:2016wjt}.

\bibitem{Caron-Huot:2020cmc}
S.~Caron-Huot and V.~Van~Duong, ``{Extremal effective field theories},''
  \href{http://dx.doi.org/10.1007/JHEP05(2021)280}{{\em JHEP} {\bfseries 05}
  (2021) 280}, \href{http://arxiv.org/abs/2011.02957}{{\ttfamily
  arXiv:2011.02957 [hep-th]}}.

\bibitem{Note4}
A priori, one can make the more general assumption that $M(s,t)$ and $\partial
  _t M(s,t)$ each exhibit level truncation at distinct sequences, $t=\xi (k)$
  and $t=\protect \tilde \xi (k)$, respectively. However, even in this case we
  can prove that enforcing these conditions implies that $\xi (k)=\protect
  \tilde \xi (k)$.

\bibitem{Bern:2008qj}
Z.~Bern, J.~J.~M. Carrasco, and H.~Johansson, ``{New relations for gauge-theory
  amplitudes},'' \href{http://dx.doi.org/10.1103/PhysRevD.78.085011}{{\em Phys.
  Rev. D} {\bfseries 78} (2008) 085011},
  \href{http://arxiv.org/abs/0805.3993}{{\ttfamily arXiv:0805.3993 [hep-ph]}}.

\bibitem{Arkani-Hamed:2003pdi}
N.~Arkani-Hamed, H.-C. Cheng, M.~A. Luty, and S.~Mukohyama, ``{Ghost
  condensation and a consistent infrared modification of gravity},''
  \href{http://dx.doi.org/10.1088/1126-6708/2004/05/074}{{\em JHEP} {\bfseries
  05} (2004) 074}, \href{http://arxiv.org/abs/hep-th/0312099}{{\ttfamily
  arXiv:hep-th/0312099}}.

\bibitem{Cheng:2006us}
H.-C. Cheng, M.~A. Luty, S.~Mukohyama, and J.~Thaler, ``{Spontaneous Lorentz
  breaking at high energies},''
  \href{http://dx.doi.org/10.1088/1126-6708/2006/05/076}{{\em JHEP} {\bfseries
  05} (2006) 076}, \href{http://arxiv.org/abs/hep-th/0603010}{{\ttfamily
  arXiv:hep-th/0603010}}.

\bibitem{Dubovsky:2006vk}
S.~L. Dubovsky and S.~M. Sibiryakov, ``{Spontaneous breaking of Lorentz
  invariance, black holes and perpetuum mobile of the 2nd kind},''
  \href{http://dx.doi.org/10.1016/j.physletb.2006.05.074}{{\em Phys. Lett. B}
  {\bfseries 638} (2006) 509},
  \href{http://arxiv.org/abs/hep-th/0603158}{{\ttfamily arXiv:hep-th/0603158}}.

\bibitem{Note5}
The graviton amplitude in general relativity is $8\pi G\protect \,{\protect
  \cal R}^4/stu$, where the expression for ${\protect \cal R}^4$ can be found
  in, e.g., Ref.~\cite {Arkani-Hamed:2022gsa}.

\bibitem{CEMZ}
X.~O. Camanho, J.~D. Edelstein, J.~Maldacena, and A.~Zhiboedov, ``{Causality
  constraints on corrections to the graviton three-point coupling},''
  \href{http://dx.doi.org/10.1007/JHEP02(2016)020}{{\em JHEP} {\bfseries 02}
  (2016) 020}, \href{http://arxiv.org/abs/1407.5597}{{\ttfamily arXiv:1407.5597
  [hep-th]}}.

\bibitem{Haring:2022cyf}
K.~H\"aring and A.~Zhiboedov, ``{Gravitational Regge bounds},''
  \href{http://dx.doi.org/10.21468/SciPostPhys.16.1.034}{{\em SciPost Phys.}
  {\bfseries 16} (2024) 034}, \href{http://arxiv.org/abs/2202.08280}{{\ttfamily
  arXiv:2202.08280 [hep-th]}}.

\bibitem{Weinberg:1965nx}
S.~Weinberg, ``{Infrared Photons and Gravitons},''
  \href{http://dx.doi.org/10.1103/PhysRev.140.B516}{{\em Phys. Rev.} {\bfseries
  140} (1965) B516}.

\bibitem{Weinberg:1964ew}
S.~Weinberg, ``{Photons and Gravitons in $S$-Matrix Theory: Derivation of
  Charge Conservation and Equality of Gravitational and Inertial Mass},''
  \href{http://dx.doi.org/10.1103/PhysRev.135.B1049}{{\em Phys. Rev.}
  {\bfseries 135} (1964) B1049}.

\bibitem{Elvang:2016qvq}
H.~Elvang, C.~R.~T. Jones, and S.~G. Naculich, ``{Soft Photon and Graviton
  Theorems in Effective Field Theory},''
  \href{http://dx.doi.org/10.1103/PhysRevLett.118.231601}{{\em Phys. Rev.
  Lett.} {\bfseries 118} (2017) 231601},
  \href{http://arxiv.org/abs/1611.07534}{{\ttfamily arXiv:1611.07534
  [hep-th]}}.

\bibitem{Note6}
We find the partial wave coefficients for $R(n,t)$, describing the amplitude
  $M$, to be given by\protect \vspace {-0.5mm} \begin {equation*}\protect
  \hspace {7.5mm} \begin {aligned} a_{n,\ell }&=\protect \genfrac {}{}{}1{\left
  (1+\protect \frac {2\ell }{D-3}\right )\Gamma \left (\protect \frac
  {D-1}{2}\right )}{\left (\protect \frac {1+3\lambda }{2}\right
  )_{n-1}^{2}}\scalebox {1.3}{\protect \raisebox {-0.35mm}{$\Sigma
  $}}_{i=0}^{n-1}\scalebox {1.3}{\protect \raisebox {-0.35mm}{$\Sigma
  $}}_{j=\ell }^{2n-2-i}\scalebox {1.3}{\protect \raisebox {-0.35mm}{$\Sigma
  $}}_{k=0}^{\lfloor (j-\ell )/2\rfloor }a_{n,\ell }^{i,j,k}\\ a_{n,\ell
  }^{i,j,k} &= \protect \genfrac {}{}{}1{(-1)^{i}(n-i)_i^2
  S_{1}(2n-2-i,j)}{i!k!(j-\ell -2k)!}\protect \genfrac {}{}{}1{j!(2-n)^{j-\ell
  -2k}(n+\lambda -1)^{\ell +2k}}{2^{j+\ell +2k}\Gamma \left (\protect \frac
  {D-1}{2}+\ell +k\right )}, \end {aligned} \end {equation*} where $S_1$ are
  unsigned Stirling numbers of the first kind. For $n\protect \,{=}\protect
  \,2$, $(2-n)^{j-\ell -2k}$ should be replaced with $\delta _{2k+\ell ,2}$.

\bibitem{Note7}
We find the partial wave coefficients for $\protect \hat R(n,t)$, describing
  the amplitude $A$, to be given by\protect \vspace {-0.5mm} \begin
  {equation*}\protect \hspace {7.5mm} \begin {aligned} \protect \hat a_{n,\ell
  } &=\protect \genfrac {}{}{}1{\left (1+\protect \frac {2\ell }{D-3}\right
  )\Gamma \left (\protect \frac {D-1}{2}\right )}{\left (\protect \frac
  {1+3\lambda }{2}\right )_{n-1}}\scalebox {1.3}{\protect \raisebox
  {-0.35mm}{$\Sigma $}}_{j=\ell }^{n-1}\scalebox {1.3}{\protect \raisebox
  {-0.35mm}{$\Sigma $}}_{k=0}^{\lfloor (j-\ell )/2\rfloor } \protect \hat
  a_{n,\ell }^{j,k} \\ \protect \hat a_{n,\ell }^{j,k}&=\protect \genfrac
  {}{}{}1{S_{1}(n-1,j)}{k!(j-\ell -2k)!}\protect \genfrac
  {}{}{}1{j!(2-n)^{j-\ell -2k}(n+\lambda -1)^{\ell +2k}}{2^{j+\ell +2k}\Gamma
  \left (\protect \frac {D-1}{2}+\ell +k\right )}. \end {aligned}\protect
  \vspace {-1.5mm} \end {equation*} For $n=2$, $(2-n)^{j-\ell -2k}$ should be
  replaced with $\delta _{\ell ,1}$.

\bibitem{Note8}
Analytic bootstraps have appeared that assume a product ansatz. However, this
  is not a physical property of scattering, but rather just the form of the
  function. In particular, a physical measurement cannot determine whether an
  amplitude is expressible in product form.

\bibitem{Berman:2024wyt}
J.~Berman and H.~Elvang, ``{Corners and Islands in the S-matrix Bootstrap of
  the Open Superstring},'' \href{http://arxiv.org/abs/2406.03543}{{\ttfamily
  arXiv:2406.03543 [hep-th]}}.

\bibitem{Caron-Huot:2016icg}
S.~Caron-Huot, Z.~Komargodski, A.~Sever, and A.~Zhiboedov, ``{Strings from
  massive higher spins: the asymptotic uniqueness of the Veneziano
  amplitude},'' \href{http://dx.doi.org/10.1007/JHEP10(2017)026}{{\em JHEP}
  {\bfseries 10} (2017) 026}, \href{http://arxiv.org/abs/1607.04253}{{\ttfamily
  arXiv:1607.04253 [hep-th]}}.

\bibitem{Berman:2023jys}
J.~Berman, H.~Elvang, and A.~Herderschee, ``{Flattening of the EFT-hedron:
  supersymmetric positivity bounds and the search for string theory},''
  \href{http://dx.doi.org/10.1007/JHEP03(2024)021}{{\em JHEP} {\bfseries 03}
  (2024) 021}, \href{http://arxiv.org/abs/2310.10729}{{\ttfamily
  arXiv:2310.10729 [hep-th]}}.

\bibitem{Chiang:2023quf}
L.-Y. Chiang, Y.-t. Huang, and H.-C. Weng, ``{Bootstrapping string theory
  EFT},'' \href{http://dx.doi.org/10.1007/JHEP05(2024)289}{{\em JHEP}
  {\bfseries 05} (2024) 289}, \href{http://arxiv.org/abs/2310.10710}{{\ttfamily
  arXiv:2310.10710 [hep-th]}}.

\bibitem{Cheung:2016wjt}
C.~Cheung and G.~N. Remmen, ``{Positivity of Curvature-Squared Corrections in
  Gravity},'' \href{http://dx.doi.org/10.1103/PhysRevLett.118.051601}{{\em
  Phys. Rev. Lett.} {\bfseries 118} (2017) 051601},
  \href{http://arxiv.org/abs/1608.02942}{{\ttfamily arXiv:1608.02942
  [hep-th]}}.

\bibitem{Arkani-Hamed:2022gsa}
N.~Arkani-Hamed, L.~Eberhardt, Y.-t. Huang, and S.~Mizera, ``{On unitarity of
  tree-level string amplitudes},''
  \href{http://dx.doi.org/10.1007/JHEP02(2022)197}{{\em JHEP} {\bfseries 02}
  (2022) 197}, \href{http://arxiv.org/abs/2201.11575}{{\ttfamily
  arXiv:2201.11575 [hep-th]}}.

\end{thebibliography}\endgroup

\end{document}